\documentclass[preprint,authoryear,12pt]{elsarticle}

\newcommand{\plin}{$P_{\rm lin}$}

\usepackage{amssymb}
\usepackage{amsmath}
\usepackage{comment}

\usepackage[driverfallback=ps2pdf,%
a4paper=true,%
breaklinks=true,%
colorlinks=true,%
pdfauthor={First Author et al.},%
pdftitle={Template for manuscripts in Advances in Space Research}%
]{hyperref}

\journal{Advances in Space Research}

\makeatletter
\def\ps@pprintTitle{%
	\def\@oddhead{\centerline{\scriptsize Accepted for publication in Adv. Space Res. doi: \url{https://doi.org/10.1016/j.asr.2022.07.034}}}
	\let\@evenhead\@empty
	\def\@oddfoot{\centerline{\scriptsize \textcopyright\ 2022.}}%
	\let\@evenfoot\@oddfoot}
\makeatother

\begin{document}

\begin{frontmatter}
	
	
	
	\title{Short timescale imaging polarimetry of geostationary satellite Thor-6: the nature of micro-glints}
	

\author[1,2,3]{Klaas Wiersema}\corref{cor1}
\cortext[cor1]{Corresponding author} \ead{k.wiersema@lancaster.ac.uk}
\author[2,4]{Paul Chote}
\author[5]{Jonathan Marchant}
\author[6]{Stefano Covino}
\author[7]{Justyn R. Maund}
\author[8]{Alexander Agathanggelou}
\author[8]{William Feline}
\author[8]{Simon George}
\author[9]{Grant Privett}
\author[1]{Brooke Simmons}
\author[5]{Iain A. Steele}

\address[1]{Physics Department, Lancaster University, Lancaster, LA1 4YB, UK}
\address[2]{Department of Physics, University of Warwick, Gibbet Hill Road, Coventry, CV4 7AL, UK}
\address[3]{School of Physics and Astronomy, University of Leicester, University Road, Leicester, LE1 7RH, UK}
\address[4]{Centre for Space Domain Awareness, University of Warwick, Gibbet Hill Road, Coventry, CV4 7AL, UK}
\address[5]{Astrophysics Research Institute, Liverpool John Moores University, Liverpool Science Park IC2, 146 Brownlow Hill,Liverpool L3 5RF, UK}
\address[6]{Istituto Nazionale di Astrofisica / Brera Astronomical Observatory, via Bianchi 46, 23807 Merate (LC), Italy}
\address[7]{Department of Physics and Astronomy, University of Sheffield, Hicks Building, Hounsfield Road Sheffield, S3 7RH, UK}
\address[8]{Defence Science and Technology Laboratory, Portsdown West, Fareham, PO17 6AD, UK}
\address[9]{Defence Science and Technology Laboratory, Porton Down, Salisbury, SP4 0JQ, UK}

\begin{abstract}
Large  constellations of orbiting communication satellites will become an important source of noise for present and future astronomical observatories. Mitigation measures rely on high quality predictive models of the position and expected brightness of these objects. Optical linear imaging polarimetry holds promise as a quantitative tool to improve our understanding of the physics of reflection of sunlight off satellite components and through which models of expected brightness can be improved. We present the first simultaneous short-timescale linear polarimetry and optical photometry  observations of a geostationary satellite, using the new MOPTOP imaging polarimeter on the 2m Liverpool Telescope. Our target, telecommunication satellite Thor-6, shows prominent short timescale glint-like features in the lightcurve, some as short as seconds. Our polarimetric observations overlap with several of these  micro-glints, and have the cadence required to resolve them. We find that the polarisation lightcurve is remarkably smooth, the short time scale glints are not seen to produce strong polarimetric features in our observation. We show how short timescale polarimetry can further constrain the properties of the components responsible for these micro-glints.
\end{abstract}

\begin{keyword}
Geosynchronous Earth Orbit\sep Optical Imaging\sep 
Polarization
\end{keyword}

\end{frontmatter}

\parindent=0.5 cm


\section{Introduction}
The characterisation of the reflection of sunlight by orbiting artificial satellites has become an increasingly important and urgent field of research in recent years, not least because of the rapid build-up of large mega-constellations of communication satellites. The reflected light of these objects is bright enough to noticeably impact sensitive astronomical observations \citep[e.g.][]{Mcdowell,Hainaut,RawlsSATCON2} at a large range of wavelengths, not just on the ground but also from low-earth orbit. 
Predictive models of both the ephemerides and the expected brightness of satellites are therefore of crucial importance to predict, evaluate and potentially mitigate their impact on sensitive astronomical observations \citep[e.g.][]{Hainaut}. Most of the current efforts have focussed on obtaining (multi-colour) broadband flux lightcurves \citep[e.g.][]{Horiuchi,TregloanReed,MrozZTF}, and basic models have been created to evaluate expected brightness as a function of sun-observer-satellite angle  \citep[e.g.][]{Hainaut,Mallamamodel,Cole,Bassa,Lawler}. 

Many satellites show glint features in their lightcurve, during which their brightness dramatically increases during a short period of time. Glints form through specular (or near-specular) reflection from relatively flat reflective  parts of the satellite (e.g. solar panels) at a specific range of sun-satellite-observer angles. Satellite glints can be mistaken for astronomical sources \citep[e.g.][]{Schaefer} and form an undesirable foreground in short timescale transient searches \citep[e.g.][]{Corbett,Karpov}.

Many of the satellites that are of greatest concern to astronomical observatories show brightnesses close to the detector saturation point of sensitive astronomical telescopes, and glinting may therefore form an additional risk factor \citep[e.g.][]{Hainaut}. The timescales of glint features are determined by the rate of change of geometry, e.g. in rotating bodies glints are very short. The shape of the reflecting features also imprints on the glint duration. Some satellites show a variety of glint timescales and amplitudes \citep[e.g.][]{Hall,Chote}. 
As shown by \cite{Vrba}, an ideal flat reflector on a geostationary orbit produces a glint that lasts around $\sim2$ minutes. Many observed glints last significantly longer than this (with timescales of around an hour), and show lower peak amplitudes than in the ideal reflector case. This indicates that the reflecting components giving rise to the glint, e.g. a solar panel, is not an ideal flat but for example consist of multiple flat pieces that are somewhat tilted with respect to one another (e.g. \citealt{Vrba}). Some geostationary satellites show glint-like features in their lightcurves with durations much shorter than traditional glints. In the following we will refer to those as micro-glints, for which we adopt a working definition of glint-like brightenings with durations below 2 minutes in geostationary orbit. 
Glints (and micro-glints) are not just a nuisance, but can also form a valuable tool to inform models of satellite reflections  \citep[e.g.][]{Hall}. Polarimetry directly diagnoses the orientation as well as the material properties of the reflecting surfaces, it can therefore solve many of the existing  degeneracies in glint models, and provide the necessary physical parameters needed for quantitative analytical modelling of reflection of satellites, both in glint phases and outside of glints.

Reflection of light off a surface induces linear polarisation. The resultant wavelength-dependent polarisation degree and polarisation angle are strong functions of the angle of reflection and the physical properties of the reflecting material. The latter are captured in the complex index of refraction $n_c$, which is defined in terms of the refractive index $n$ and the extinction coefficient $k$  as $n_c = n - i*k$; the linear polarisation induced by specular reflection will be maximal at the  Brewster angle of the reflecting material. 
Satellites consist of several reflecting surfaces, with different relative orientations and with different refractive indices $n_c$. The main reflecting surfaces are the solar panels, the side(s) of the spacecraft bus that faces the observer (which may be covered in multi-layer insulation, MLI), and the antenna dishes. As a satellite orbits the Earth, the angle of the sunlight reflecting of different elements rapidly changes:  we should therefore see changing polarisation degree and angle as a function of time. When the reflection angle gets close to the Brewster angle of the material of a reflecting component, we may expect a strong change in the total observed polarisation. The polarisation properties of spacecraft materials have been studied numerically and in the lab \citep[e.g.][]{Pasqual,Beamer,Peltoniemi}.  
In principle, the problem can be reversed, and the satellite's orientation and physical parameters of reflection can be empirically determined from well-sampled multi-colour polarimetric lightcurves (polarisation degree and angle), assuming some basic shape properties and geometry \citep[aided by  lightcurve analysis, e.g.][]{Seo}  as priors, by fitting a Mueller matrix chain (describing the optical action of each reflecting element) directly onto the total observed wavelength-dependent polarisation as a function of angle. This is a method frequently used in calibration and design of optical telescopes and instruments, where we can fit for the orientation and indices of refraction of reflecting surfaces as free parameters in the components of the Mueller matrix chain made up of all optical components \citep[see e.g.][for an example]{EFOSC2calib}. To do this successfully for satellites requires multi-wavelength, high cadence, high  accuracy (low systematic errors) polarimetry over a substantial range of solar phase angles (the Sun-object-observer angle). Such datasets are not yet publicly available. However, single wavelength, lower cadence polarimetry datasets are an  important first step, to identify the main satellite components responsible for the observed polarisation  \citep[e.g.][]{Speicher,Beamer,Kosaka}, to provide an inventory of empirical polarisation behaviour for a variety of satellite platforms \citep[e.g.][]{Speicher} and to postulate a sensible range of priors for more quantitative fitting methods.

Glints are particularly helpful lightcurve features  \citep[e.g.][]{Vrba}, as these are expected  to show substantial  amounts of optical linear polarisation \citep{Speicher,Zimmerman}. They are bright, which reduces  statistical errors of polarisation measurements. While some polarimetric data exists of glints \citep[e.g.][]{Zimmerman}, short duration events like micro-glints are not well studied polarimetrically to date. \cite{Speicher} have shown indications of optical polarimetric signals associated with micro-glints, but their study  was limited to relatively long lasting micro-glints (several minutes) studied at relatively poor temporal resolution, with  generally only one or two polarimetric datapoints covering the lightcurve feature. To use micro-glints as a quantitative tool, we need polarimetry at timescales of seconds, with small polarimetric uncertainties ($\sigma_P\lesssim0.2\%$). 

The data discussed in this paper were taken as part of a 
pilot programme to use a new imaging polarimeter \citep[MOPTOP, the  Multi-colour OPTimised Optical Polarimeter;][]{Jermak2016,Jermak2018, ShresthaMOPTOP} on the robotic  Liverpool Telescope \citep{SteeleLT} to study changes in orientation of satellites through their polarisation signatures, particularly the docking of the MEV-2 vehicle with the geostationary Intelsat 10-02 satellite. During that programme, we observed another geostationary satellite, Thor-6, as a calibration observation (i.e. a secondary calibrator): that observation is the topic of this paper.  This object was selected because of its close proximity on the sky to the MEV-2 + Intelsat 10-02 pair, and its well monitored lightcurves (Chote et al.~in prep.). Thor-6 is also interesting in its own right: this satellite shows bright and frequent micro-glints in its optical lightcurves, and therefore enables a first search for polarimetric signals of micro-glints at timescales of a few seconds in reflected optical light of geostationary satellites. 
While geostationary satellites generally do not pose a risk to astronomical observations (in contrast to satellite constellations at lower orbits), they are a useful testbed for the type of observational studies required to better understand the reflection properties of satellites that do pose a risk but are  more challenging to study, e.g. because of their rapid movement on the sky (beyond the maximum non-sidereal tracking speed of many older $\gtrsim2$m class telescopes). 

In this paper we show our acquisition, analysis and calibration of the MOPTOP polarimetry of Thor-6, as an example of the capabilities of MOPTOP for short timescale optical polarimetry of moving objects, and compare the data to simultaneous lightcurves. We show how our data, and future data covering a larger range of time, can be used to place constraints on (or measure directly) the nature of the structural components of the satellite causing  micro-glints.

\section{Thor-6 and polarimetry of geostationary  satellites}
Thor-6 (also known as Intelsat 1W) is a currently active geostationary telecommunication satellite, primarily providing television broadcasting services.   It is owned by Telenor Satellite Broadcasting AS, 
 and built by Thales Alenia Space. It was launched on 29 October 2009 by an Ariane 5ECA launch vehicle, from Kourou, French Guyana. Thor-6 uses the Thales Alenia Spacebus-4000B2 platform. Its shape is broadly of the ``box-wing'' type: a box-shaped bus, several large dish antennas and two long rectangular solar panels extending from the north and south faces of the bus, spanning a few tens of meters.

Geostationary satellites have  been studied using optical polarimetry before. These observations were generally performed at relatively low cadence. \cite{Speicher} observed a small sample of geostationary objects using a polarimeter on a small telescope, finding 
a relatively large diversity in polarisation lightcurves, likely reflecting a diversity in satellite  shape and geometry. The authors used an instrument that recorded two channels, a horizontally and a vertically polarised component. Based on changes of the relative strength of the horizontal and vertical components as a function of viewing geometry, some inferences can be made to which satellite  component is contributing most polarised light. \cite{Zimmerman} observed a small sample of geostationary satellites with a small telescope, using quasi-simultaneous polarimetric and low resolution spectroscopic observations, finding evidence for an increase in linear polarisation during times that a glint was visible in the lightcurve.
For a satellite in low-Earth orbit we expect similar polarimetric behaviours (after correcting for orbit orientation differences: geostationary satellites are located in a fairly narrow equatorial belt, whereas low earth orbit satellites cover a wide range of inclinations), with the key difference that they traverse the range of solar phase angles over a much shorter timespan, compressing the relevant timescales, which makes obtaining diagnostic data more challenging.  

\cite{Kosaka} observed geostationary satellite Express-AM5 using a much  larger telescope, the 2m Nayuta telescope, and a polarimeter, for $\sim5$ hours at a cadence of 90 seconds, forming one of the highest quality and highest  cadence  polarimetric datasets of a geostationary satellite to date. In  their data, they see a minimum in the optical polarisation (\plin$\sim1\%$) around the time of the minimum phase angle, with rapidly increasing linear polarisation after the minimum phase angle (with values increasing up to \plin$\sim14\%$ in the phase angle interval covered by their observations). The measurements from \cite{Kosaka} provide full Stokes $Q,U,I$ (see Section \ref{sec:moptopcalib}) and are calibrated onto the absolute polarisation degree and angle values system \citep[][where polarisation angle towards North is $0^{\circ}$ and East is $90^{\circ}$]{transactionsIAU_1973}. However, at 90 second cadence, rotation of the satellite with respect to the observer can be significant and cause artefacts in the data;  observations at higher time resolution are needed to avoid these.

\begin{figure*}[ht]
\centering
	\includegraphics[width=8cm]{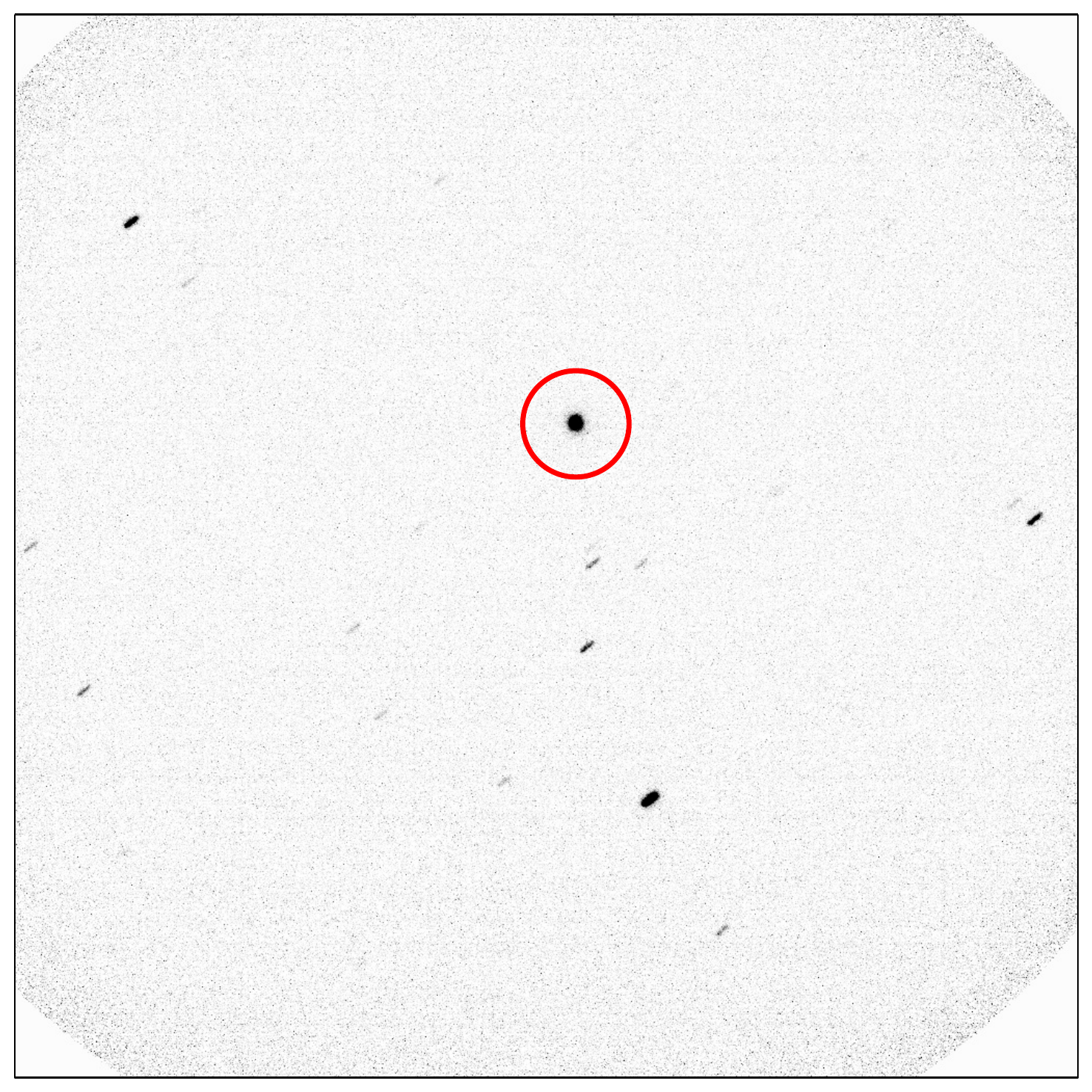}
    \caption{A full, representative, single MOPTOP image from our observation, with Thor-6 circled. This is image 1\_e\_20210504\_14\_26\_4, i.e. an image of cam1, with run number 14, rotation number 26 and waveplate position 4. The integration time for this image is 0.4 seconds (the fixed value for {\tt FAST} mode observations). Stars can be seen as streaks in the image. Because of the short integration time of individual images, these streaks are relatively short.}
    \label{fig:singleimage}
\end{figure*}

\section{Observations}
The observations reported in this paper consist of a 800 second high-cadence imaging polarimetry observation taken with the MOPTOP instrument on the Liverpool Telescope, and a high-cadence optical lightcurve taken simultaneous with the polarimetry, from the same geographical location, using the University of Warwick test telescope. 
\subsection{MOPTOP observations}
The polarimetric observation of Thor-6 in this paper was   
performed robotically by the 2m  Liverpool Telescope (LT; \citealt{SteeleLT}), located on the island of La Palma, Spain; under proposal number DL21A02 (PI Wiersema).  We used the  
Multi-colour OPTimised Optical Polarimeter (MOPTOP) imaging polarimeter \citep{Jermak2016,Jermak2018, ShresthaMOPTOP}. This dual beam polarimeter, optimised for time-domain astrophysics, uses a continuously rotating half-wave plate and a wiregrid polarising beamsplitter;  two scientific CMOS cameras (Andor Zyla sCMOS cameras) record the images of the two orthogonally polarised beams simultaneously, hereafter we refer to these two cameras as {\em cam1} and {\em cam2}. The detector readouts are synchronised to the waveplate rotation. Sixteen images are recorded by each camera for every full (360 degree) waveplate rotation;  for details and design motivation see \cite{Jermak2016,Jermak2018} and \cite{ShresthaMOPTOP}. A total of 32 images are therefore recorded for each full waveplate rotation, at mean waveplate angles of $0^\circ, 22.5^\circ, .. , 337.5^\circ$.
%
%
MOPTOP can be used with two fixed wave plate rotation speeds, the {\tt SLOW} and {\tt FAST} mode. In the former, the rotation period of the wave plate is 80 seconds, in the latter 8 seconds. 
This translates to a frame exposure time of 4.0 s in {\tt SLOW} mode, and 0.4 s in {\tt FAST} mode (the remaining time is used for read-out). For the observation discussed in this paper we used the {\tt FAST} rotator observing mode, which is best suited to bright sources and provides good time resolution. The MOPTOP observations of Thor-6 used a $R$ filter (MOP-R), covering the wavelength range $\sim580-695$ nm. 

The observation of Thor-6 was prepared and executed as follows: in the afternoon before the night of observation, we retrieved the most recent Two-Line Elements (TLEs) for the target from the Celestrak website (\url{https://www.celestrak.com/NORAD/elements/}). 
We then generated an ephemeris table for the geographical location and altitude of the Liverpool Telescope (LT) using the JPL Horizons On-Line Ephemeris System, 
with a time resolution of 1 minute. 
Within the LT phase2 tool, the target was uploaded as an ephemeris table (a so-called {\em ephemeris target}), and observations were defined using the {\tt FIXED} observing mode, i.e. a fixed time was defined at which the observations were to be started, within a user-configurable tolerance  (the so-called {\em slack},  which we set at 10 minutes for this observation). The resulting predefined observation was entered into the LT queue, with no constraints placed on the seeing and sky brightness; observations were selected, scheduled and executed robotically. 
When executing a given ephemeris target observation, the telescope will interpolate between the coordinates given in the ephemeris file for acquisition and tracking. 
Note that LT can not auto-guide on moving objects. The observation was taken with the Cassegrain mount angle rotation set to zero degrees. 
The airmass for this observation was 1.275; the first exposure was started at 03:19:08.511 UT on 5 May 2021, and we observed for a total on-target time of 800 seconds (100 wave plate rotations; this timespan is currently the limit for a single FAST mode observation). The weather conditions were good and the seeing at the start of the polarimetric observation was $\sim1"$. The solar declination at the start of the observation was $+16.207$ degrees.

\subsection{Photometric observations}\label{sec:lightcurve}

\begin{figure*}
\centering
	\includegraphics[width=13cm]{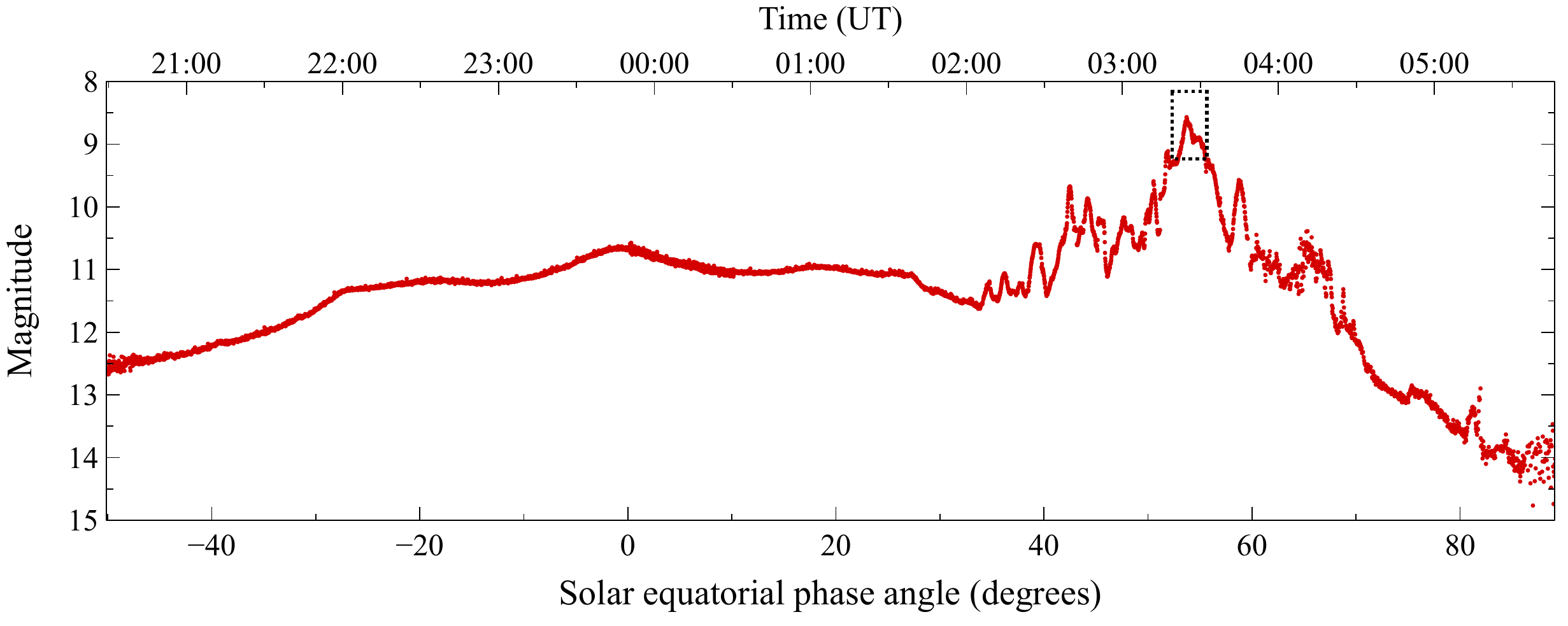}
    \caption{Optical lightcurve of Thor-6 in the night starting 4 May 2021 (see Section \ref{sec:lightcurve}). Both the solar equatorial phase angle (defined as the longitudinal component of the angle between the satellite and the anti-solar point;  \citealt{Payne2007}) and the time of observation (UT) are shown on the horizontal axes. The box marks the timespan of the MOPTOP observations, shown in detail in Figure \ref{fig:pola}. Magnitudes are in the Vega system, calibrated onto Gaia $G$ band values for field stars (Chote et al.~in prep.).}
    \label{fig:lightcurve}
\end{figure*}

We obtained a large number of high-cadence optical light curves between February and May 2021 as part of an observation campaign studying the rendezvous, proximity operations, and docking of MEV-2 with Intelsat 10-02 \citep[see][]{George}. 
Observations were made using the University of Warwick's test telescope, also located on La Palma, which was configured for these observations using a Takahashi Epsilon 180ED wide-field astrograph with an Andor Marana sCMOS detector. This combination provided a $2.6^\circ \times 2.6^\circ$ field of view with a pixel scale of 4.5"/pixel. Simultaneous full-night light curves were obtained for Intelsat 10-02, MEV-2, Thor-5, Thor-6, and Thor-7, which are located close together on the sky, all within the field of view of this telescope. We used cadences between 1 s and 5.5 s (the shortest exposures were necessary to avoid saturation during the main glint features around local midnight, where brightness could peak as high as 5$^\text{th}$ magnitude). The full observation campaign and data reduction procedures will be discussed in a future publication (Chote et al.~in prep); in this paper we will use the lightcurve coinciding with the MOPTOP polarimetric observations, which is shown in Figure~\ref{fig:lightcurve}. The photometry is calibrated by integrating over the streaks of an ensemble of suitable calibration stars (selected to avoid blending with other star streaks, of suitable brightness, and non-variable) and matching the instrumental magnitude against Gaia to obtain a zero point in Gaia G and a colour term that is evaluated at (G\textsubscript{BP} - G\textsubscript{RP})$_\odot$ = 0.82, the Gaia colour of the Sun \citep{Casagrande2018}. Typically around 400 calibration stars are used per image. The light curve (Figure \ref{fig:lightcurve}) is plotted as a function of the Solar equatorial phase angle, which is defined as the longitudinal component of the angle between the satellite and the anti-solar point \citep{Payne2007}.

\section{MOPTOP data reduction and analysis}\label{sec:moptopcalib}
The MOPTOP data reduction procedure  is  described by \cite{ShresthaMOPTOP} and at the MOPTOP website. 
The raw frames undergo bias and dark subtraction, and are corrected using a flatfield constructed from a stack of flatfield images at all 16 waveplate positions  \citep[i.e. the flatfield is the same for the images at all 16 waveplate positions, see][]{ShresthaMOPTOP}. This method works well for dual beam polarimeters under certain conditions  \citep[for a discussion see e.g.][]{Patat}. 

\begin{figure*}[ht]
\centering
	\includegraphics[width=8cm]{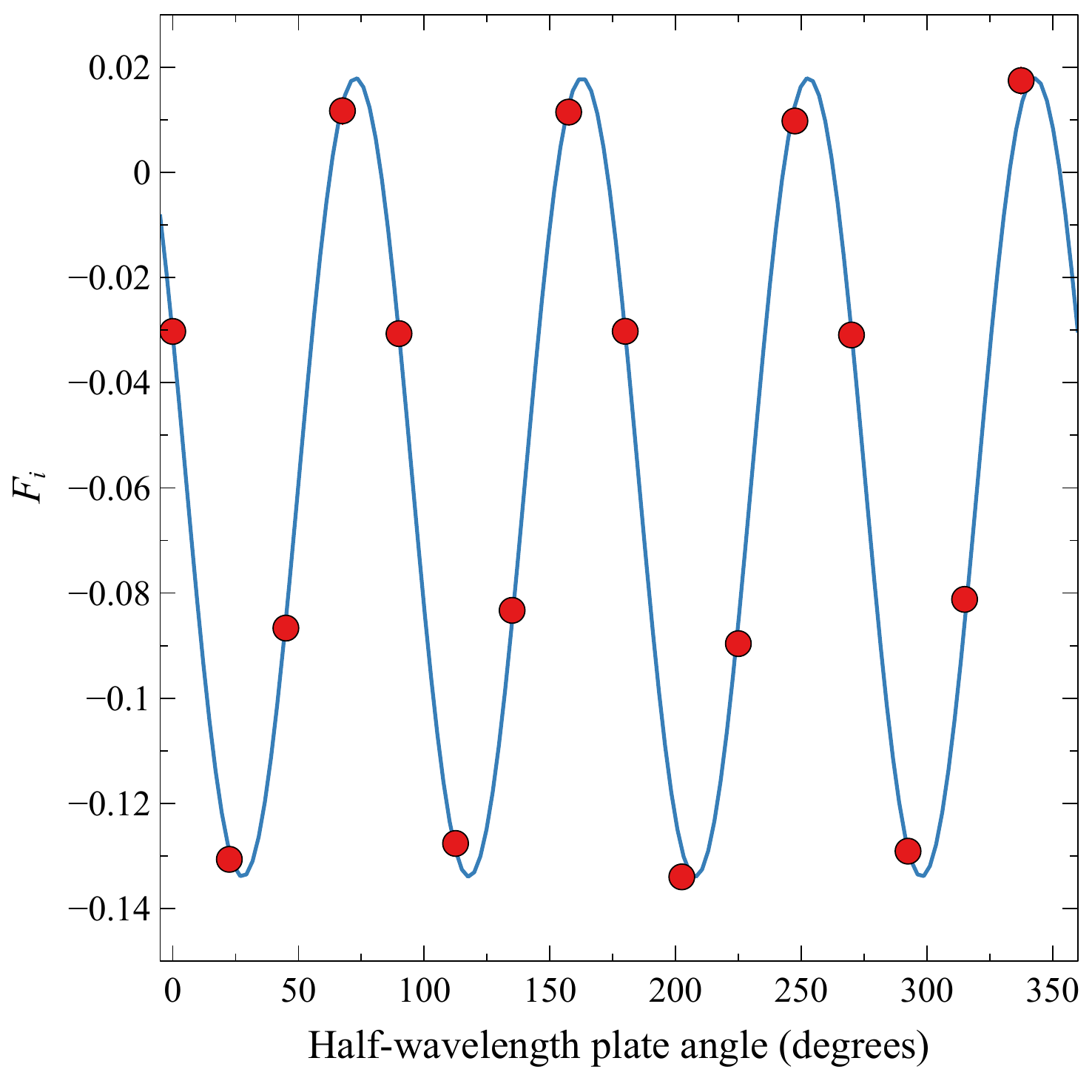}
    \caption{The red symbols show the normalised flux differences $F_i$ of one of our observations (a single full waveplate rotation) of a polarised standard star with MOPTOP in {\tt FAST} mode, using the MOP-R filter. The blue solid line is the sum of the $n=0$ and $n=4$ Fourier components (Section \ref{sec:moptopcalib}). }
    \label{fig:fouriermodulation}
\end{figure*}

Our analysis procedure of the reduced data (i.e. measuring fluxes and computing polarisation) differs slightly from the methods set out in \cite{ShresthaMOPTOP}, and we detail our approach in the following. First, the reduced data are sorted by date and epoch and some basic properties of the data are retrieved from the file headers. The centroid of the target is then measured using the {\tt IRAF} {\em starfind} and {\em imcentroid} tasks. As geostationary objects maintain a broadly constant altitude and azimuth, stars in the field move rapidly over the detector, forming streaks (Fig \ref{fig:singleimage}), which in rare cases may influence the centroiding when they happen to pass very close to the target. We use fairly strict sigma-clipping values in the centroiding procedure to eliminate this effect; our target is very bright. We measure the fluxes of the target in the {\em cam1} and {\em cam2} images, using aperture photometry in {\tt IRAF}, using the {\em apphot} package. The aperture radius is chosen as 2 times the average FWHM (full width at half maximum) of a Gaussian fit to the object point spread function (the target is unresolved), and is kept fixed for all exposures: the seeing was stable during the MOPTOP observation duration to within 0.1 arcsecond. Aperture radii are the same for {\em cam1} and {\em cam2}. An annulus shaped region was used to determine the local sky background level. Hereafter we use the notation  $f_{{\rm cam1},i}$ and $f_{\rm cam2,i}$ for the target flux in camera 1 and camera 2 at the $i$-th waveplate angle. 
We compute normalised flux differences $F_i = (f_{{\rm cam1},i} - f_{{\rm cam2},i}) / (f_{{\rm cam1},i} + f_{{\rm cam2},i})$ for each exposure set at each angle  $\phi_i$ of the half-wave plate.

First, we analyse a set of polarised standard stars (three observations of HD 155197, one of Hiltner 960 and one of VI Cyg 12), all observed in {\tt FAST} mode and using the MOP-R filter (these stars span the magnitude range 10.6-9.4 mag; standard sidereal tracking is used for these observations). We measure their fluxes in the same way as for Thor-6, and compute their normalised flux differences $F_i$. 
We then perform a simple Fourier analysis on the standard star $F_i$ values to verify the modulation behaviour of MOPTOP, following \cite{PatatCAFOS}, using the expression  \citep{Fendt,PatatCAFOS}:
\[
F_i = a_0 + \sum_{n=1}^{N/2}\left[ a_n \cos(n(2{\pi}i/N)) + b_n \sin(n(2{\pi}i/N))\right],
\]
where $a_n, b_n$ are the Fourier coefficients, $N$ the number of waveplate angles, and $i$ the $i$-th angle as above. As explained in \cite{PatatCAFOS}, an ideal dual beam polarimeter of the design of MOPTOP would have all its Fourier power in the $n=4$ component, and all other components would be zero. We fit the $F_i$ data of the polarised standard stars using this Fourier prescription, using the {\it symfit}  package \citep{Symfit} in Python. 
As expected, we find that the only statistically significant terms (found with $\gtrsim5\sigma$ significance) are the $n=0$ (i.e. $a_0$) and the $n=4$ terms.  Pleochroism ($n=2$ component) is not significantly detected. Figure \ref{fig:fouriermodulation} shows an example MOPTOP MOP-R band {\tt FAST} mode dataset of polarised  standard star Hiltner 960 (observed on 8 May 2021), where a model consisting only of the $n=0$ and $n=4$ terms is shown to provide an excellent description of the data. 
We describe the polarisation state of incoming light through the Stokes vector $\vec{S} = (I,Q,U,V)$; note that some authors prefer the equivalent notation $\vec{S} = (S_0,S_1,S_2,S_3)$ for the Stokes vector components. In the following we will not consider the Stokes $V$ (or $S_3$) component:  in reflection scenarios as we consider here, optical circular polarisation is mainly caused by cross-talk, i.e. circular polarisation is induced when the reflected light was somewhat linearly polarised before reflection, so there is cross-talk between the $Q,U$ and $V$ Stokes parameters. This can for example take place in scenarios where light gets reflected twice, or in reflection from complex layered materials. We therefore generally expect low values of circular polarisation, and in the following we focus on the linear polarisation.

Given the result of the Fourier analysis above, we use a simple prescription for calculating the Stokes parameters as: 
\begin{align}
\begin{split} \label{eq:q}
q = Q/I = \frac{2}{N} \sum\limits_{i=0}^{N-1}F_{i}{\rm cos}\left(\frac{i\pi}{2}\right)
\end{split} \\
\begin{split} \label{eq:u}
u = U/I = \frac{2}{N} \sum\limits_{i=0}^{N-1}F_{i}{\rm sin}\left(\frac{i\pi}{2}\right)
\end{split}
\end{align}
for each set of 4 waveplate angles, i.e. we compute four independent values for $q,u$ for each full waveplate rotation. In other words, waveplate angles 0$^\circ$, 22.5$^\circ$, 45$^\circ$ and 67.5$^\circ$ give $q_1,u_1$; 90$^\circ$, 112.5$^\circ$, 135$^\circ$ and 157.5$^\circ$ give $q_2,u_2$; etc. We use these independent measurements as our individual datapoints, giving a time resolution of 2 seconds.
The errors on the Stokes parameters are calculated through standard error propagation. The values are corrected for instrumental polarisation using the values for the MOP-R band listed on the MOPTOP website ($q_{\rm inst} = +0.0091,u_{\rm inst}=-0.0302$ for MOP-R), 
which we verified using a MOPTOP dataset of an unpolarised standard star taken close in time to the Thor-6 observation. We compute the linear polarisation \plin\ and the polarisation angle $\theta$ via 
\begin{align}
\begin{split} \label{eq:p}
P_{\rm lin} = \sqrt{q^{2} + u^{2}}
\end{split} \\
\begin{split} \label{eq:theta}
\theta = \frac{1}{2}{\rm arctan}\left(\frac{q}{u}\right), 
\end{split} 
\end{align}
where the quadrant-preserving arctan is used. 
In the conversion from $q,u$ to $P_{\rm lin},\theta$ we  expect to encounter the effects of polarisation bias  \citep{Serkowski1958,Wardle1974,Simmons1985}. This bias arises from the fact that $q$ and $u$ can be positive or negative, with their errors generally following a Normal distribution. In contrast, $P_{\rm lin}$ is positive definite (equation 3), and has a Ricean probability distribution.  In the presence of noise on $q$ and $u$, we can therefore over-estimate $P_{\rm lin}$ in situations with low signal to noise, this is often referred to as polarisation bias. There are a large number of studies offering various correction techniques to take this into account. We use the modified asymptotic (MAS) estimator, as defined in \cite{Plaszczynski2014} to correct for polarisation bias, but find that in all observations of Thor-6 the polarisation bias plays no significant role (this is not surprising: the flux signal to noise $f/\sigma_f$ is very high, as is the polarisation signal to noise $P/\sigma_P$).  The resulting polarisation values are corrected for instrumental de-polarisation, for which we use the multiplicative value tabulated on the MOPTOP website for the MOP-R band, as derived from observations of polarised standard stars, which we verified using the observations of polarised standard stars mentioned above. The final calibration step consists of placing the polarisation angle in the correct absolute frame, for which we follow the prescription from the MOPTOP website: $\theta_{\rm true} = \theta_{\rm inst} + \theta_{\rm rotskypa} + c$, where $\theta_{\rm inst}$ is the instrumental polarisation angle found above, $\theta_{\rm rotskypa}$ is the instrument rotation angle as tabulated in the {\em rotskypa} header keyword, and $c$ is a constant offset.  We use the polarised standard star observations to compute the average offset between instrumental polarisation angle (corrected for their $\theta_{\rm rotskypa}$ values) and their values from the literature: we used the values tabulated in \cite{Schmidt} for HD\,155197, Hiltner 960 and VI Cyg 12. Note that Hiltner 960 and VI Cyg 12 may show some signs of variability \citep{Blinov}. We find a $1.0^{\circ}$ systematic error on the absolute angle values for our set of standard star observations. This calibration should place the polarisation on the IAU definition of polarisation angle \citep{transactionsIAU_1973}. 

\begin{figure*}
	\includegraphics[width=13cm]{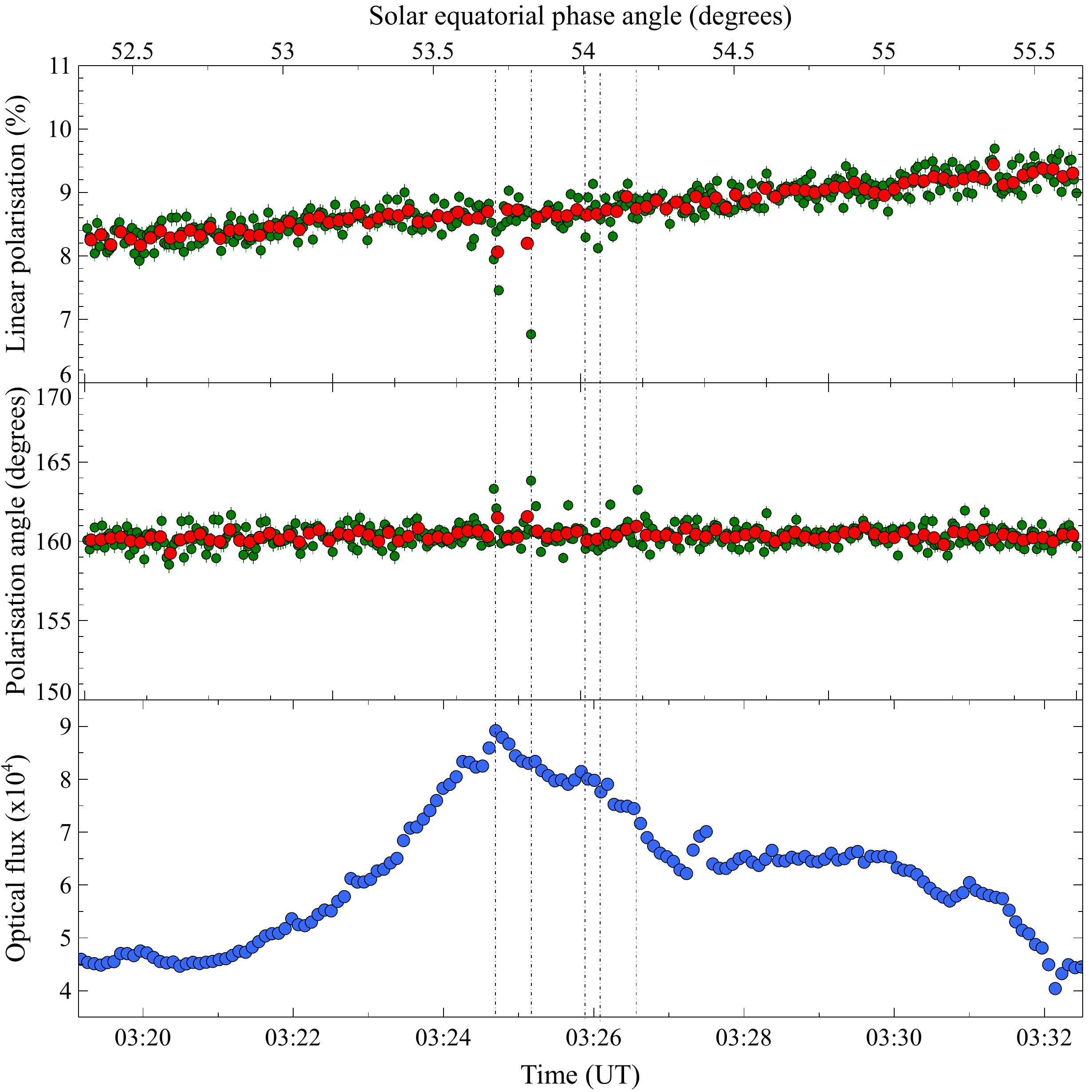}
    \caption{The polarisation lightcurve from the 800s MOPTOP observation described in this paper, with the linear polarisation degree ($P_{\rm lin}$) in the top panel, and the calibrated polarisation angle in the middle panel. Green points are the individual, independent measurements, red points are 4-point binned values. The bottom panel shows the optical lightcurve during the same time interval, here shown in linear flux values (in analog-to-digital units, ADU) as measured by the Warwick test telescope (see section \ref{sec:lightcurve}) rather than magnitudes (Fig. \ref{fig:lightcurve}), to allow a more intuitive comparison with the linear polarimetry. Several lightcurve features are clearly visible at short and longer timescales, with a range of amplitudes. The thin dashed vertical lines indicate the position of some of the outlier polarisation datapoints. 
    The plot symbol size is larger than the formal errorbars for the lightcurve data and the binned polarimetry. }
    \label{fig:pola}
\end{figure*}

The final polarisation lightcurve, of both linear polarisation degree and angle, with the individual 2 second datapoints and a 4-point binned average (8 seconds), is shown in Figure \ref{fig:pola}.  

\section{Discussion}\label{sec:discussion} 
\subsection{Polarimetry of moving objects  with LT+MOPTOP}\label{sec:discMOPTOP}
The polarisation lightcurve of Thor-6 (Figure \ref{fig:pola}) shows that polarimetry at short timescales of moving objects with magnitudes typically seen for active  geostationary satellites is indeed feasible with the MOPTOP instrument on LT.
The field of view of MOPTOP ($\sim7'\times7'$) is large enough to reliably place a moving object in the field of view, if the object has a relatively recent TLE. Our observation in this paper, and those of MEV-2 (Wiersema et al.~in prep) show that in most cases,  moving targets can be placed close to the optical axis by the robotically operated LT, where the instrumental polarisation is well calibrated \citep{ShresthaMOPTOP}. Observations of much  faster moving objects, such as the Starlink satellites in low-earth orbit, are more challenging for LT, as their angular velocity over the sky exceeds the current limits of the telescope (J.~Marchant, priv.~comm.).
Trailed imaging polarimetry (where the satellite creates trails in the images) may be possible in some cases, but at the expense of significantly increased systematic errors. 
This tracking speed limit is not a limitation for some other observatories and commonly used mounts, and a short-timescale imaging polarimetry campaign is important to better inform efforts to mitigate against the impact of mega-constellations on astronomical observations. Another important property is the magnitude of the satellite: brighter objects allow the use of shorter exposures and waveplate rotation timescales for the polarimeter for a given $\sigma_P$ requirement. In many cases, observations at larger phase angles hold important diagnostic power, but satellites are generally fainter then (Figure \ref{fig:lightcurve}); telescopes of $\sim2$m class play an important role to provide accurate high-cadence polarimetry in those cases.
Thor-6 was magnitude $\sim9.4-8.6$ during the interval covered by MOPTOP. This gave good statistical errors  \citep[of order $0.1\%$, i.e. similar to the MOPTOP systematic errors,][]{ShresthaMOPTOP}. Even at the peaks of the observed glint signatures, the peak of the target point spread function was relatively far from image saturation or non-linearity limits, indicating that somewhat brighter glints can still be safely observed by MOPTOP in {\tt FAST} mode. 

\begin{figure}[ht]
\centering
	\includegraphics[width=8cm]{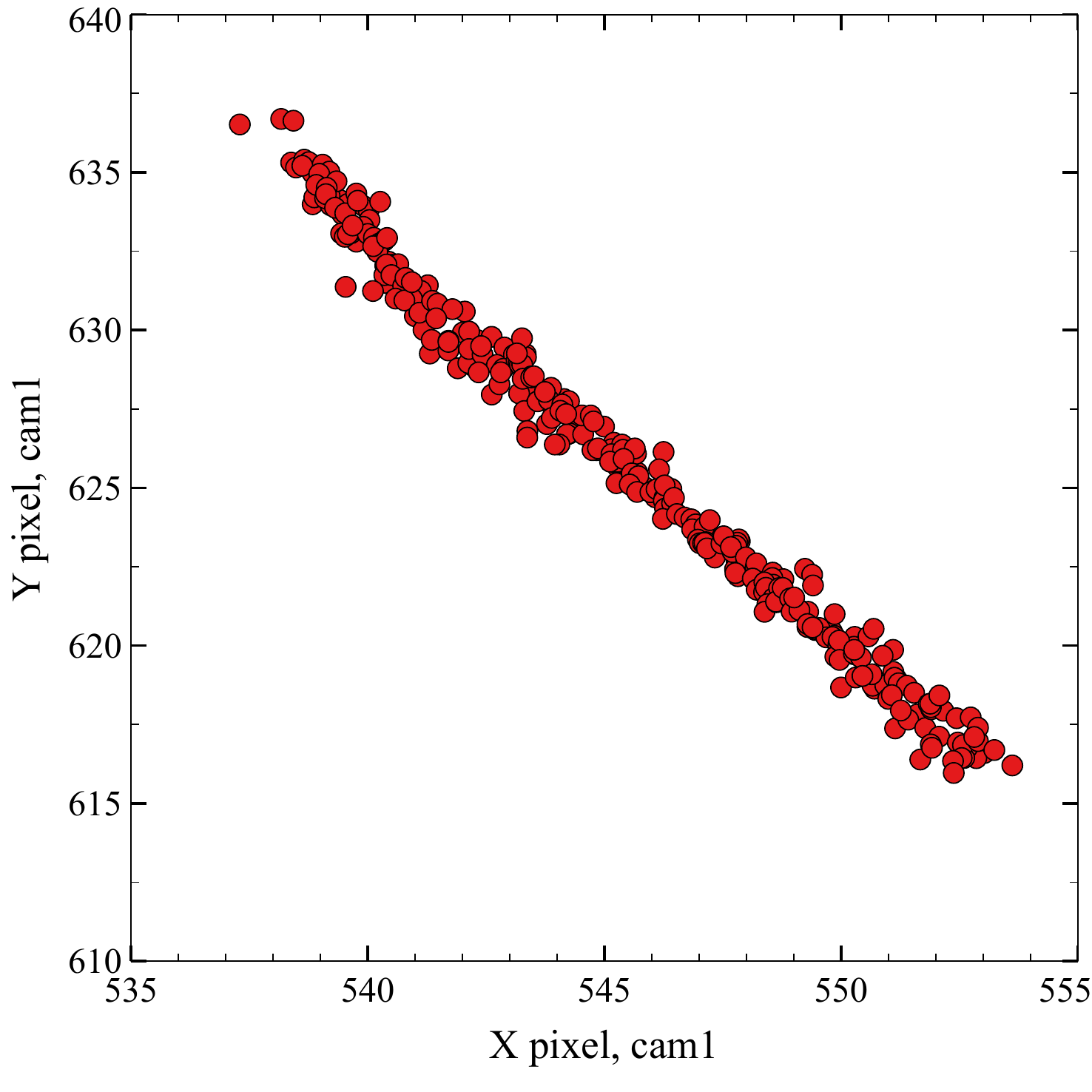}
    \caption{The centroid pixel position of Thor-6 in the {\em cam1} data. The position gradually and monotonically  drifts from top left to bottom right in this diagram. This is almost entirely along the North-South direction (specifically, moving northwards). The  pixel scale of MOPTOP is 0.42"/pixel, the total drift is $\sim10.8$". }
    \label{fig:driftxycam1}
\end{figure}

\begin{figure}
	\includegraphics[width=\columnwidth]{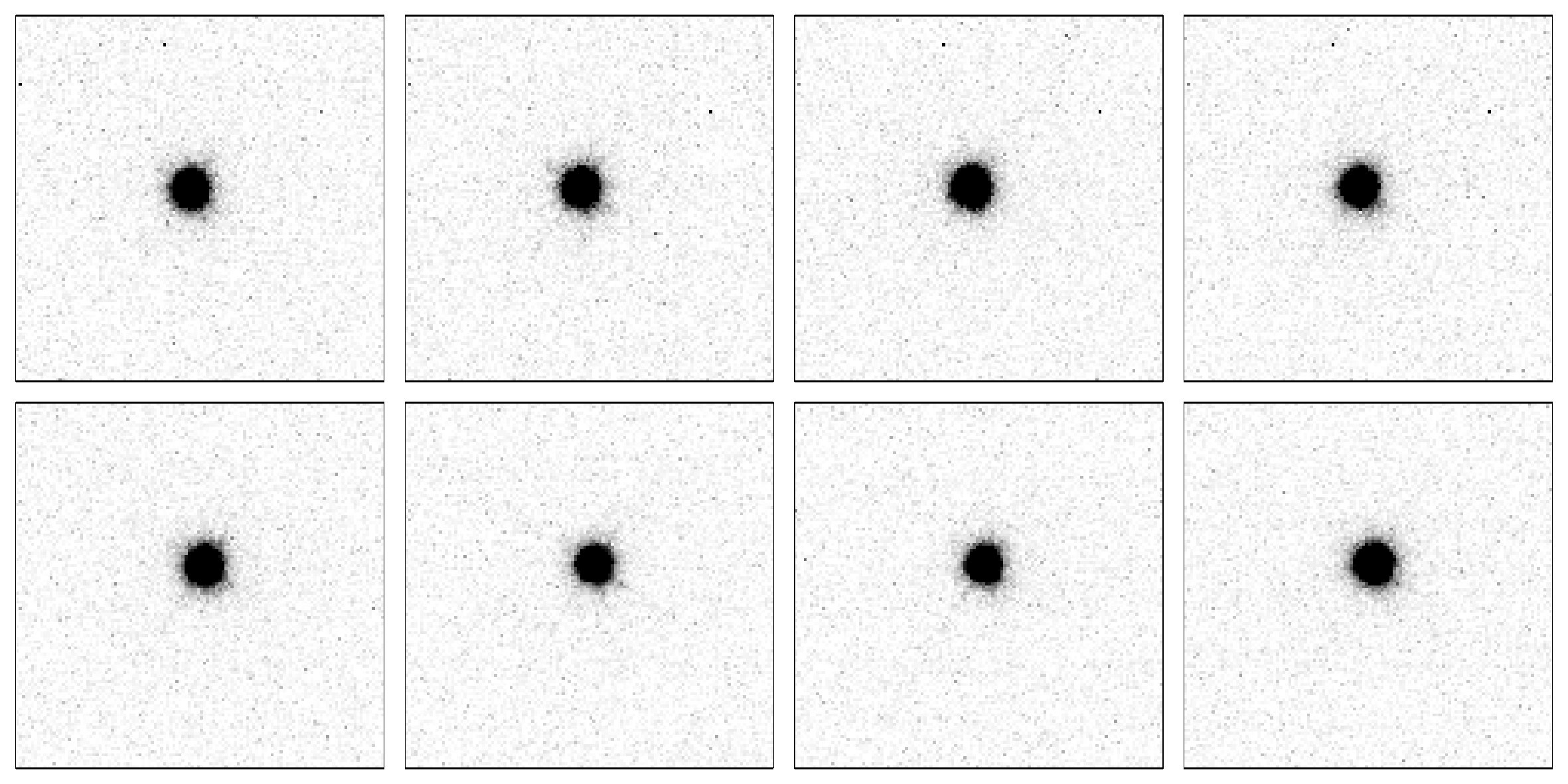}
    \caption{Shown are the eight images belonging to the datapoint with unexpectedly low $P_{\rm lin}$ at phase angle 53.72 degrees (see Figure \ref{fig:pola}), as an example of a outlier datapoint. Shown are $120\times120$ pixel cut-outs ($50.4\times50.4$ arcsec), with on the top row the four {\em cam2} images and on the bottom row {\em cam1}. There is no clear signature of a background star passing over the object. }
    \label{fig:outlier}
\end{figure}

The scatter of the datapoints around the general trend in Figure \ref{fig:pola} is somewhat larger than one might expect based on the formal statistical errors of the individual datapoints, indicating some non-optimal effects play a role. One of those is the drift of the  target over the detector: in an ideal polarimeter, the target would always occupy the same pixels, so that the beam-swapping that takes place by using four waveplate angles (equations \ref{eq:q} and \ref{eq:u}) minimizes the  effects of imperfect flatfielding, and so that hot pixels and other defects can be more efficiently corrected for. In our dataset, we see some drift of the target over the image during the observation. Figure \ref{fig:driftxycam1} shows the centroid of Thor-6 move in a fairly monotonic fashion in X and Y pixel coordinates, mostly along the North-South direction, moving Northwards. The total position change in the 800 second observation is approximately 10.8 arcseconds. Comparison with the calibrated astrometry from the Warwick test telescope (Section \ref{sec:lightcurve}) shows that this drift is primarily caused by errors in the TLE orbit prediction, rather than faults in the telescope tracking (LT  can not auto-guide on moving objects). 
At the timescale of the individual sets of 4 waveplate angles  that make up one $q,u$ measurement (2 seconds) the drift is negligible, and the target point spread function is well described by a Gaussian profile throughout.
 Another possible reason for additional scatter (and potentially a fraction of the outlier datapoints) is the rapid passage of field stars through the source aperture or the sky annulus region (see Figure \ref{fig:singleimage}). Observatories with poorer resolution (large effective pixel scales) suffer this effect more than ones with better resolution. The spatial resolution of MOPTOP is excellent (0.42" per pixel, seeing of 1"), so this will only affect a very small number of datapoints. Visual inspection of the exposures confirms that this is indeed not the cause of the outlier datapoints (see Figure \ref{fig:outlier} for an example), the aperture and annulus radii are relatively small and  the satellite did not cross through dense star fields in our observing time. 
Finally, the sCMOS detectors used on MOPTOP show ``popcorn'' noise \citep[see][]{ShresthaMOPTOP}: random telegraph noise appearing as hot pixels at random locations in each frame. These may alter flux measurements somewhat when appearing by  chance in the source aperture. We use four waveplate angles for each single $q,u$ pair measurement, this beam-swapping reduces the influence of such single pixel noise events in single images. Combining more than four angles to make one $q,u$ measurement further reduces this influence, but this comes at the cost of temporal resolution. In Figure \ref{fig:pola} we therefore show the single (2 sec) datapoints and a binned version averaging four datapoints to one point.  

\subsection{Micro-glints and Thor-6}
The MOPTOP polarimetric lightcurve  of Thor-6 (Figure \ref{fig:pola}) shows a relatively smooth
trend, with a broadly linear increase in polarisation degree \plin\ from $\sim8.2$\% to $\sim9.3$\% in the 800 s covered by our observation. 
During this interval, the polarisation angle $\theta$ stays broadly constant. On top of this long-timescale trend only some low-amplitude wiggles may be present, limited to low polarimetric amplitudes ($\lesssim0.3$\%) and short timescales (tens of seconds at  most) on top of the general trend. 
The slowly increasing polarisation and the slow evolution of the polarisation angle in the MOPTOP interval is reminiscent of the polarisation curve of the  geostationary satellite Express-AM5, presented by \cite{Kosaka} (their Figure 2), which had very similar \plin\ and slowly varying polarisation angle at the phase angle of the MOPTOP observations of Thor-6 ($\sim 52.32-55.66$ degrees, Fig \ref{fig:pola}). The polarimetric lightcurve in \cite{Kosaka} is sampled at much lower temporal resolution (90 seconds vs the MOPTOP 2 seconds), but is of longer duration. The authors attribute the majority of the behaviour of \plin\ in their dataset of Express-AM5 to reflection of the solar panels of the satellite, with possible smaller contributions from the bus and/or the antenna dishes. This is based on the behaviour of the polarisation angle with  time and the value of the linear polarisation as a function of reflection angle compared to the values found in the lab by \cite{Beamer}. It is important to point out that the Express-AM5 satellite has a different platform from Thor-6, though shares many of the main features, e.g a box-like bus, large antennas and large extended wing-like solar panels. We consider it likely that the longer timescale trend in the MOPTOP polarisation data of Thor-6 is similarly caused by reflection of the solar panels, using the same arguments as made by \cite{Kosaka}. 

Thor-6 is an interesting target because of the presence of short duration, bright, glint-like flares in the high cadence lighcurves (Chote et al.~in prep) on top of the smoother diffuse reflection. Additional multi-filter observations (Chote et al.~in prep) showed that there were no significant colour changes associated with these features. This means we can reliably compare features in our wide-band lightcurves with the $R$ band polarimetry.  Figure \ref{fig:lightcurve} shows the lightcurve at the night of the MOPTOP observations  (i.e. this lightcurve was taken simultaneous with the polarimetry, with a telescope at nearly the same geographical location). Up until $\sim02$ UT  the lightcurve is smooth, showing only broad features, with a peak near zero phase angle, which is commonly seen in lightcurves of geostationary satellites. After $\sim02$ UT a phase of rapid lightcurve variability starts, with many short duration, overlapping,  glint-like peaks, some only barely resolved at the cadence of our lightcurve observations. Two broader features, at around $\sim45^\circ$ and $\sim54^\circ$ solar phase angle seem present, with many short peaks superposed on them. 
These short glints, which we call micro-glints here, appear to have a wide distribution of amplitude, duration and shape, with some lasting considerably shorter than a minute.  

The lower panel of Figure \ref{fig:lightcurve} shows the  small portion of the lightcurve which covers the time interval of the MOPTOP polarimetry. In Figure \ref{fig:pola} we show that same lightcurve in instrumental flux units (analogue to digital units, ADU) rather than magnitudes, together with the polarisation lightcurve, to allow easy comparison; the MOPTOP data cover several micro-glints with a  range of amplitudes and timescales. It is important to note here that the LT and the Warwick telescope are on the same mountain peak, separated by just $\sim250$ 
meters. This is smaller than  the expected size of the glint patch striking the Earth for an ideal flat reflector 
\citep{Vrba}. We can therefore directly compare the lightcurve and the polarisation. 

Firstly we note that some outlier datapoints (or regions of larger scatter) are visible in the unbinned polarisation lightcurves. A subset of these may be attributable to some instrumental noise effects  (Section \ref{sec:discMOPTOP}), but it is clear that they take place near the peaks of the highest amplitude micro-glints, which may indicate a causal relation. 
There is also some indication that some more gradual changes/ripples (spanning $\sim10-30$ sec) in the polarisation degree (with amplitudes $\sim0.3\%$) occur at the time of some of the micro-glints, e.g. near the peak of the prominent micro-glint at $\sim53.7^\circ$ phase angle ($\sim337$ sec in Figure \ref{fig:pola}). However, several other micro-glints seen in the flux lightcurve  seem not to have  produced a detectable polarimetric feature, for example the one at $\sim54.4^\circ$ phase angle. 

The duration and amplitude of glint features from an ideal flat reflector is given by the crossing time of the sun spot size at the observer \citep[e.g.][]{Vrba}. In many cases, glints in (not rapidly rotating) geostationary satellites appear to take much longer than this, which is generally attributed to a non-ideal reflector, e.g. a panel made up of smaller facets that are  not perfectly aligned  \citep[e.g.][]{Vrba}. 
\cite{Zimmerman} observed the optical linear polarisation of a small sample of geostationary 
satellites during regular (relatively long lasting) glints, quasi-simultaneous with low resolution spectroscopic observations. In their data the authors observe that several satellites show a polarimetric signature around the glint, as expected from specular reflection off relatively large surfaces \citep[e.g.][]{Vrba}, generally showing an increase in linear polarisation (tens of percent). In some other objects \cite{Zimmerman} did not detect such behaviour. Our MOPTOP data has much better time resolution and sensitivity, allowing us to  detect  very small polarisation changes ($\sim0.2\%$) on short timescales. This enables us to search for similar effects in micro-glints. Reflection off smaller satellite parts can in principle generate lower amplitude small glints; a distinguishing signature is how the faint glints and micro-glints behave over multiple nights as a function of solar phase angle \citep{Hall} and as a function of wavelength. As mentioned above, polarimetry can be an independent diagnostic. One possibility for the origin of the micro-glints is reflection of small reflecting components of the spacecraft bus or the antennas. Large sections of the bus are covered in multi-layer insulation (MLI), which can reflect highly specularly  \citep[e.g.][]{Peltoniemi,Rodriguez}, and may therefore give strong polarisation signatures under favourable reflection angles. A simplified laboratory setup has indeed shown strong polarisation features using a square bus model with Kapton  (a polyimide film frequently used in MLI) as an example MLI \citep{Beamer}, with strong polarisation spikes near reflection angles close to the phase angle of our observation  \citep[i.e. the broad peaks in the lightcurve, e.g. at $\sim54^\circ$,  are reminiscent of the features seen in the analysis by][]{Beamer}. 

Given the above it is somewhat surprising to only see weak evidence for polarisation spikes associated with the micro-glints.  
In some satellites, the MLI layer is fairly taut and smooth, in others it is more ``wrinkly''. In the latter case, many individual reflecting facets/sections of MLI may contribute to the received light of the bus. As the phase angle changes, small sections may glint briefly, not unlike a disco ball. For the polarisation we expect the largest source of reflected light (the  solar panels) to dominate the observed polarisation parameters at relatively large values of the phase angle, when the solar panel polarisation  is high \citep[e.g.][]{Kosaka}. On top of this baseline, the short glints from the MLI facets will give short polarisation spikes (as well as flux spikes), whose polarisation degree and angle depend on the material properties.  In this wrinkly MLI scenario, there may be many superposed (micro-)glints and reflections (as seems supported by the flux lightcurve in Figure \ref{fig:lightcurve}), both diffuse and specular, whose polarimetric components sum up - but as the polarisation angles differ, this sum may result in a less obvious polarimetric signature at a given time. If the facets of MLI giving rise to the observed polarisation are fairly small compared to the size of the reflecting side of the satellite (and the other sources of polarised light, e.g. the solar panels, are large), we may expect the polarimetric signatures of the micro-glints to be relatively low in amplitude.  In addition we note that at even with our high time resolution of 2 seconds, we may suffer from a degree of smearing of the signal. However since  most of the micro-glints have resolved rise and fall times in the flux lightcurves, which have somewhat lower cadence than the  MOPTOP sampling, this seems likely to be a relatively small effect.  

In the MOPTOP interval we presented here (just 800 sec of data), the number of isolated, well characterisable micro-glints is relatively small. Future short timescale observations covering a much  larger number of micro-glints are important to increase our sensitivity through statistics, and allow meaningful correlation studies. Observations covering a large phase angle range will be important, not  just for the benefit of the modelling of the micro-glints but also to quantitatively model the dominant underlying components, such  as the  solar panels. 
Thor-6 is sufficently bright over an entire night (Figure \ref{fig:lightcurve}) that {\tt FAST} mode observations are suitable over the entire night, i.e. fast timescale polarimetry is possible also at high phase angles. In addition, the datapoints obtained in {\tt FAST} mode can be adaptively binned to decrease polarimetric errors at the expense of time resolution, allowing high accuracy measurements for somewhat fainter objects as well.  We are also somewhat helped by the fact that the linear polarisation increases at increasing phase angle (\citealt{Kosaka}).  At the brighter end, objects brighter than $\sim5-6$ mag may saturate using {\tt FAST} mode.

\subsection{Future Prospects}
Thor-6 is unresolved in our MOPTOP images \citep[as expected;][]{Hart}. Some satellites in low and medium earth orbits will be resolvable by MOPTOP on the LT (or a similar instrument and telescope combination; a 5m satellite at 500km altitude can span $\sim2$"): the  pixel scale of MOPTOP is 0.42", and good and stable seeing conditions are common at La Palma. For these objects, imaging polarimetry during glints would yield a particularly rich amount of information, as the glint features can be directly attributed to specific sections of the spacecraft, removing some free parameters in a quantitative (e.g.~Mueller matrix chain) modelling. As these objects move rapidly  over the sky, demands on tracking speeds are much higher than for geostationary objects, but within reach of many modern telescope mounts.

Another route of future  progress is the use of simultaneous multi-wavelength polarimetry, as the polarisation signal from (specular) reflection by a given material is strongly wavelength dependent. While spectro-polarimetry would provide the most ideal dataset, this would require either a large telescope or the use of long exposure times to obtain data with sufficiently small statistical errors per wavelength bin. Combined with inevitable overheads of most existing instruments (e.g. CCD readout, waveplate rotation) and the need to accurately place and retain a target in the spectrograph slit, this makes it challenging to spectro-polarimetrically study the time resolved properties of micro-glints. Broadband imaging polarimetry is an easier option.  MOPTOP currently has a single arm \citep{ShresthaMOPTOP}, and multi-colour data can therefore only be taken consecutively, through filter changes using the filter wheel. Future MOPTOP upgrades envisage the use of more than one arm, with light split by  dichroic elements  \citep[as is done by the DIPol-UF imaging polarimeter for example,][]{Piirola}, which would enable strictly simultaneous multi-colour imaging polarimetry and greatly increase the science yield in the field of satellite observations. This is of particular interest for the proposed New Robotic Telescope (NRT), a robotic successor to LT with a primary mirror size of $\sim4$m, for which time-domain polarimetry is a key priority, and whose light collecting power would enable high-speed polarimetry of fainter satellites. 

A third interesting possibility is observing the same satellite with two (or more) widely separated  (in longitude or latitude) telescopes with polarimeters at the same time, using the same wavelength (filter).  Of particular interest are situations when the telescopes are separated by distances of order the sun 
spot size of the (micro-)glints on the ground  \citep[e.g.][]{Vrba}, which would add a powerful diagnostic to identify the exact reflecting component responsible for the glinting behaviour, and its orientation. At large latitude differences, the viewing angles onto the reflecting areas orthogonal to the east-west direction is different enough that the resulting integrated polarisation outside of glints will be noticeable different. Given that the viewing angle offsets can be precisely calculated, this will provide a powerful additional constraint on polarisation model inversion fits. A first attempt to do this combining the LT (with MOPTOP) with the University of Leicester 0.5m telescope (with the LE2Pol 
optical dual-beam imaging  polarimeter; Wiersema et al.~in prep) was unsuccesful because of local COVID-19 access restrictions. 

We finally point out that Thor-6 has been in space for a relatively long time (Section 2). The effects of the solar wind and intense  ultraviolet radiation environment on the reflecting components of satellites is not well understood. The MLI and solar panels of the satellite may have aged considerably in this environment, with significant changes in their reflection properties compared to laboratory measurements. Future polarimetric observations of Thor-6, and polarimetric observations (at the same phase angles) of other Thor satellites with different times in orbit  may help to diagnose the effects of aging.

\section{Conclusions}
In this paper we present a single 800 second observation of optical imaging polarimetry in the $R$ band of geostationary satellite Thor-6, obtained with the MOPTOP instrument mounted on the 2m robotic Liverpool Telescope at La Palma (Spain). Our data probe short timescales (down to 2 seconds) at high polarimetric accuracy. 
We add to that dataset a high cadence optical lightcurve from the University of Warwick’s test telescope at La Palma, obtained simultaneously to the polarimetry.  The lightcurve shows a large number of short timescale, high amplitude glints, often overlapping, which we refer to as micro-glints in this paper. This combined dataset is one of the most sensitive and highest cadence polarimetric observations of a geostationary  satellite to date; the polarimetric observations overlap with a period of intense micro-glinting. In our MOPTOP data, the observed linear polarisation as a function of solar phase angle is dominated by a gradual evolution, which we may ascribe to reflection off the large solar panels of Thor-6. We can exclude strong polarisation features associated with micro-glints covered by  our observation, but some faint features (bumps with a polarisation  amplitude of a few tenths of percent) may  be present in the lightcurves \plin\ and polarisation angle. In particular, some increased scatter in the polarisation data is visible at the times of some of the micro-glint peaks. To establish correlation requires a future larger sample of micro-glints observed using high cadence polarimetry, and a greatly increased phase angle coverage. Our observation shows that the robotic LT with the MOPTOP instrument is a highly suitable combination to do this.

\section*{Acknowledgments} 
We thank the anonymous referees for their constructive feedback and comments. 
The Liverpool Telescope is operated on the island of La Palma by Liverpool John Moores University
in the Spanish Observatorio del Roque de los Muchachos of the Instituto de Astrofisica de Canarias with financial support from the UK Science and Technology Facilities Council (STFC).
This work has made use of data obtained using the Warwick CMOS test telescope operated on the island of La Palma by the University of Warwick in the Spanish Observatory del Roque de los Muchachos of the Instituto de Astrofisica de Canarias.
KW acknowledges that this project has received funding from the European Research Council (ERC) under the European Union Horizon 2020 research and innovation programme (grant agreement no 725246, PI prof.~A. Levan) and support through Royal Society Research Grant RG170230 (PI dr.~R. Starling). BS and KW acknowledge support through a UK Research and Innovation Future Leaders Fellowship (MR/T044136/1; PI dr.~B. Simmons). KW thanks A.~Berdyugin for discussions about NOT DIPol-UF observations of fast moving satellites. SC acknowledges partial funding from Agenzia Spaziale Italiana-Istituto Nazionale di Astrofisica grant I/004/11/5.

\end{document}